\documentclass[12pt,preprint]{aastex}

\shorttitle{Mid-IR Polarimetry of NGC1068}
\shortauthors{Packham et al.}

\begin{document}


\title{Gemini Mid-IR Polarimetry of NGC1068: Polarized Structures
Around the Nucleus}


\author{C. Packham\altaffilmark{1}, S. Young\altaffilmark{2}, 
S. Fisher\altaffilmark{3}, K. Volk\altaffilmark{3}, R. Mason\altaffilmark{3}, 
J. H. Hough\altaffilmark{2}
P. F. Roche\altaffilmark{4}
M. Elitzur\altaffilmark{5}, 
J. Radomski\altaffilmark{6},
and E. Perlman\altaffilmark{7}}

\email{packham@astro.ufl.edu}


\altaffiltext{1}{University of Florida, Department of Astronomy, 211 Bryant 
Space Science Center, P.O. Box 112055, Gainesville, FL, 32611-2055, USA}
\altaffiltext{2}{Center for Astrophysics Research, 
University of Hertfordshire, Hatfield, AL10 9AB, UK}
\altaffiltext{3}{Gemini Observatory, Northern Operations Center, 
670 N. A'ohuku Place, Hilo, Hawaii,96720, USA}
\altaffiltext{4}{University of Oxford, Department of Astrophysics, 
Keble Road, Oxford, OX1 3RH, UK}
\altaffiltext{5}{University of Kentucky, Department of Physics and Astronomy,
600 Rose Street, Lexington, KY, 40506, USA}
\altaffiltext{6}{Gemini Observatory, Southern Operations Center, 
c/o AURA, INC., Casilla 603, La Serena, Chile}
\altaffiltext{7}{Physics and Space Sciences Department, Florida Institute of 
Technology, 150 West University Boulevard, FL, USA}


\begin{abstract}
We present diffraction limited, 10$\mu$m imaging polarimetry data for
the central regions of the archetypal Seyfert AGN, NGC1068.  The
position angle of polarization is consistent with three dominant
polarizing mechanisms.  We identify three distinct regions of polarization:
(a) north
of the nucleus, arising from aligned dust in the NLR, 
(b) south, east and west of the nucleus, consistent with dust being channeled
toward the central engine and (c) a central minimum of polarization
consistent with a compact ($\le$22pc) torus.  These observations 
provide continuity
between the geometrically and optically thick torus and the host
galaxy's nuclear environments.  These images
represent the first published mid-IR polarimetry from an 8-m class 
telescope and illustrate the potential of such observations.
\end{abstract}





\section{Introduction}
\label{intro}

The unified theory of Seyfert (Sy) type active galactic nuclei (AGN) 
holds that all types of Sy AGN are essentially the same object, viewed
from different lines of sight (LOS).  Surrounding the central engine is a
geometrically and optically thick, dusty, molecular torus, obscuring
the broad emission line region from some LOS.  In this scheme, the
Sy classification depends solely on the LOS and exact torus properties.
Such theories received a major boost through the detection of scattered,
and hence polarized, broad emission lines in the spectrum of NGC1068 
\citep{ant85}, revealing an obscured Sy 1 central engine
in the previously classified Sy 2 AGN, entirely consistent with unified 
theories. 

Whilst fundamental to unified theories, the torus remains difficult
to image directly at optical/IR wavelengths, with perhaps the most direct
observation of the torus made by speckle interferometry in the near-IR
\citep{weig04}.  Strong evidence for significant amounts 
of obscuring material in the central 100pc-scale nuclear regions of 
NGC1068, possibly in the 
form of a torus, is provided by observations of CO and HCN 
emission \citep{plan91,jack93,schin00}, and recent Chandra X-ray 
observations \citep[e.g.][]{ogle03}.  Mid-IR spectroscopy reveals 
a moderately deep ($\tau_{9.7}\approx 0.4$) silicate absorption 
feature at the nucleus \citep{roche84,sieb04}, whose strength is 
approximately constant up to $\sim$1\arcsec\ south of the brightest 
mid-IR point \citep{mason06,rhee06}. Applying the \citet{nen02} clumpy torus 
model, \citet{mason06} suggested the torus is compact ($\le15$pc), in 
good agreement with mid-IR interferometric observations \citep{jaffe04}.
Further evidence for a compact torus was found through AO fed H$_2$ 1-0S(1) 
observations \citep{davies06}, finding a 15pc clump of H$_2$ extending
from the nucleus at the same PA as the line of masers.
The observed extent of the torus, or nuclear obscuring material in NGC1068,
is partly dependent on the wavelength and/or observational technique.  
\citet{young96} used
imaging polarimetry to observe the silhouette of obscuring material
against the southern ionization cone, which they attributed to the torus, 
with a derived diameter of $\sim$200pc in the H-band.

The close proximity (1$\arcsec$ $\equiv$ 72pc) 
and high brightness of NGC1068 makes it an ideal target
for polarimetry, a traditionally photon-starved application.  Near-IR
studies of NGC1068 by \citet{pack97}, \citet{lums99} and \citet{simp02} 
clearly detected the bi-conical ionization structure in scattered light.  In
the nuclear regions, there is a trend to a position angle (PA) of polarization
being perpendicular to the radio jet with increasing wavelength.  Modeling 
of the nuclear regions requires both an extended area of scattering particles
as well as dichroic absorption of nuclear emission, possibly by dust
in, or associated with, the torus (\citet{young95}, \citet{watan03}).

\citet{bail88} found that the PA of polarization rotates by
$\sim$70$\degr$ between 4 and 5$\mu$m.  The 10$\mu$m spectropolarimetry of
\citet{ait84} showed a similar PA of polarization to that at 5$\mu$m,
and a constant degree of polarization through the silicate absorption feature.
These results are entirely consistent with the predicted 
90$\degr$ change from dichroic absorption to dichroic emission from 
aligned dust grains.  That the PA change was only $\sim$70$\degr$ is 
attributable to dilution of the dichroic emission component by 
polarized flux in 
the extended scattering cones, most likely from dichroic emission
from dust in the narrow emission line region (NLR) \citep{bail88}.

To investigate the contributions of the various polarizing mechanisms 
and structures in the nucleus of NGC1068, \citet{lums99} performed  
imaging polarimetry using a broad-band 8-13~$\mu$m filter. These data 
represented the first and only published mid-IR imaging polarimetry of 
an AGN, but their interpretation was complicated by the $\sim$0.7\arcsec\ 
resolution of the data. To take advantage of the improved spatial 
resolution attainable from an 8 m-class telescope, we obtained new 
mid-IR imaging polarimetry of NGC1068 during commissioning of this 
mode at the Gemini North telescope.

\section{Observations}

We obtained imaging polarimetry of NGC1068 during commissioning of the
polarimetry unit of Michelle \citep{glass97} on UT 2005 December
19 on the Gemini North 8.1m telescope.  These observations 
were primarily aimed at measuring the degree and position angle (PA) of 
polarization with NGC~1068 as a test object, and hence used a
limited on-source time of 148 seconds.  Michelle uses a Raytheon 320 x 
240 pixel Si:As IBC array, providing a plate scale of 0.1$\arcsec$ 
per pixel in imaging mode. Images were obtained in the 9.7$\mu$m 
($\delta$$\lambda$ = 1.0$\mu$m, 50$\%$ cut-on/off) filter only, 
using the standard chop-nod technique to remove time-variable sky
background, telescope thermal emission and so-called ``1/f"
detector noise.  The chop throw was 15$\arcsec$ and the telescope
was nodded every $\sim$90 s.  The chop throw was fixed at
0$\degr$ (N-S).  Conditions were photometric and the observations
were diffraction limited ($\sim$0.30\arcsec FWHM).

Michelle employs a warm, rotatable half wave retarder (or half wave plate, 
HWP) to modulate the polarization signal, located upstream of the entrance
window of the dewar.  A cold wire grid polarizer is used as the
polarimetric analyzer, located in a collimated beam.
Images were taken at four HWP PAs in the following sequence:
0$\degr$, 45$\degr$, 45$\degr$, 0$\degr$, 22.5$\degr$, 67.5$\degr$,
67.5$\degr$, 22.5$\degr$ in the first nod position, and the sequence 
repeated in the second nod position.  In this manner, the Stokes 
parameters can be computed as close in time as possible,
reducing the effects of variations in sky transmission and emission.  This
sequence, however, requires many motions of the HWP, and is therefore
under evaluation with a view to reducing the number of HWP motions 
to increase observing efficiency.  Data were
reduced using the Gemini IRAF package in conjunction with Starlink
POLPACK software \citep{berry03}.  The difference for each chop pair 
in a given nod
position and HWP PA was calculated, and then differenced with the
second nod position at the same HWP PA.  Images were aligned through 
shifting by fractional pixel values to account for slight image drift 
between frames, and then the Stokes parameters I, Q and U computed using
POLPACK.  A total of 20 nod positions were recorded, and residual 
array/electronic noise was removed through use of a median-filter 
noise mask.  The data were reduced through creation of four
individual I, Q and U maps and also through coadding all frames at their
respective HWP PA first and then producing a single I, Q and U map.
The S/N in the latter method is slightly higher in the individual 
Q and U maps, presumably due to a 'smoothing' of the Q and U during
the co-addition; these are the data used in this paper. 

The efficiency and zero angle calibration of the polarimeter were 
measured through
observations of two polarized sources and comparison with 
measurements published
by \citet{smith00}.  The instrumental polarization was estimated to be
$\le$0.3\% through observations of two stars that fulfilled
the criteria of (a) high proper motion (hence nearby), (b) high galactic 
latitude (to minimize the presence of Galactic dust) and (c) an
intermediate spectral type star (to minimize intrinsic stellar nebulosity). 


\section{Results}

Figure 1 shows the total flux image (color-scale and contours) with the 
polarization vectors overlaid.  The polarization vectors are plotted where
the S$/$N is $\ge$54 in total flux, and contours are linearly spaced in 
intensity, starting at a S$/$N of 27.
Figure 2 shows the polarized intensity map, produced by multiplying the
degree of polarization by the total intensity image.  As in Figure 1,
only where the S/N in the total intensity image is $\ge$54
are data plotted.  The resultant polarization vectors are contained 
within an approximate N-S oriented ellipse, major/minor axes 
1.7$\arcsec$/1.2$\arcsec$ respectively.
The integrated degree of polarization within that ellipse is 
2.48$\pm$0.57\% at a PA of 26.7$\pm$15.3$\degr$.  The errors
in the degree and PA of polarization are estimated through independent
measurements of the four individual polarization maps and computing
the standard deviation.  It should be noted that the exposure time in those
four individual maps was very short, where systematic effects could
dominate, and hence the quoted errors may be an overestimation.
The degree of polarization is higher than the 1.30\% measured 
by \citet{lums99} in a 
2\arcsec\ aperture, consistent with an increased observed 
polarization as often arises with improved 
spatial resolution.  The PA of polarization is significantly
different from Lumsden's measurement of 49$\degr$ in a 2$\arcsec$ aperture.
However, our data shows a PA rotation of 94$\degr$ from the K$_n$ band
data of \citet{pack97} and \citet{lums99}, entirely as 
expected if the dominant polarizing mechanism changes from dichroic 
absorption to emission between the two wavebands, as described 
in \S\ref{intro}.  We speculate the \citet{lums99}
$\sim$0.68$\arcsec$ results suffered significantly greater contamination in 
their beam, possibly from surrounding extended polarization, 
as compared to our $\sim$0.30$\arcsec$ results.  Additionally, 
the wider bandwidth of 
\citet{lums99} would have been more affected by the different and competing
polarizing components.

The degree and PA of polarization suggests contributions from three 
components. The first extends $\sim$1$\arcsec$ north of the 
mid-IR peak and is coincident with the inner regions of the radio jet,
with a PA of polarization approximately N-S.  
The second region extends south, east and west of the nucleus, with 
a PA of polarization of $\sim$35$\degr$.
Finally, the degree of polarization drops to a minimum very close 
to the mid-IR total flux peak, which we believe is most likely to 
arise from an unresolved polarization contribution with a PA of
polarization approximately orthogonal
to that of the more extended emission, leading to a reduction
in the measured polarization.
The polarized intensity image reveals polarized 
emission extending
north of the mid-IR peak, and two areas of enhanced flux east and west
of the mid-IR peak, and a minimum close to the mid-IR total flux peak.
Table 1 summarizes the locations and polarization components.


\section{Discussion}

Polarization at mid-IR wavelengths most likely arises from either
dichroic absorption or emission, both due to dust grains with a preferred
alignment.
The integrated PA of polarization in these data confirms and enhances the 
interpretation of the near-90$\degr$ PA flip from near- to mid-IR wavelengths,
with the factor $\sim$2.5 increase in spatial resolution providing
a more accurate and consistent result.  \citet{galli03,galli05} suggested, 
based on spatial coincidence, the [OIII] clouds in the ionization cone are 
the dominant 
mid-IR sources away from the compact torus.  The polarized flux image
shows a similar spatial correspondence with the [OIII], and 
the PA of polarization north of the nucleus is consistent with the 
interpretation of
\citet{bail88} of dichroic emission in the NLR, possibly through dust
alignment via jet streaming or a helical magnetic field 
associated with the jet.  
We discount directly observed synchrotron radiation from the radio 
jet accounting for the polarization, as an extrapolation of the 
radio emission to the mid-IR provides too little flux.
Hence, this data confirms the extended 
mid-IR polarized emission north of the nucleus is dominantly from dust in 
the ionization cone.

South, east and west of the nucleus, as the PA of polarization is 
perpendicular to that in the near-IR where the 
polarization is thought to be produced by dichroic absorption,
we suggest the dominant polarizing mechanism is dichroic emission by
grains aligned to the same field direction as the absorbing grains,
in agreement with other authors (i.e. \citet{bail88}).  Due to
the Barnett effect (\citet{laz03}, and references therein), grains align
with their short axes parallel to the local magnetic field, and the 
PA of polarization is parallel to the direction of the magnetic 
field for dichroic absorption and orthogonal for dichroic emission.
The location of the polarized
emission areas and PA of polarization is suggestive of warm aligned 
dust grains being channeled from the host galaxy toward the torus.
Indeed, the PA of the polarized flux is coincident
with the H$_2$ material that \citet{davies06} associated with
molecular material in a compact torus.

Dichroic absorption in an unresolved optically thick central region 
could account for the minimum in polarization close to the peak of 
mid-IR flux, with a PA of polarization approximately orthogonal to the more 
extended dichroic emission to the east, west and south. 
Alternatively the mid-IR flux in the innermost regions could arise from
a strong mid-IR, intrinsically unpolarized, source.
However, there is tentative 
evidence of the central regions showing a twist in the
PA of polarization, tending toward a similar PA found in the dichroic
absorption at near-IR wavelengths (e.g. \citet{pack97}), which supports 
the dichroic absorption interpretation.  In both possibilities, a
compact ($\le$0.3$\arcsec$ ($\le$22pc) diameter) torus could 
account for this result.
If correct, the polarization minimum indicates the true position of the
central engine, which is not coincident with the mid-IR total flux peak,
but displaced by $\sim$0.2$\arcsec$ to the west.

%
CO \citep{schin00} and optical HST 
\citep{catch97} observations are interpreted as evidence of a warped molecular 
disk on 100pc scales, 
partially obscuring the nuclear regions of the host galaxy and 
ionization cone pointing 
away from Earth.  Indeed, \citet{schin00} speculate this material, 
rather than a compact torus, is responsible for obscuring the AGN.
\citet{elitzur06} suggest the 100pc molecular structure is the extension
of the pc scale disk of masers (\citet{green97}, \citet{galli01}, 
\citet{galli03}).
We suggest that our data provide
continuity between the geometrically thick torus (height/radius $\sim$1)
to the flatter, larger galactic disk (height/radius $\sim$0.15).

The western polarized feature is considerably larger than the compact 
($\le$ few pc) torus suggested by several authors 
\citep[e.g.][]{jaffe04,mason06, pack05,radomski06} on the basis of 
mid-IR imaging and modeling, but much smaller than the 
suggested torus detected by \citet{young96}.  However, the feature
is detected in polarized flux, a technique that increases
contrast by removing the dominant, unpolarized, emission. 
Distinct from total flux, polarimetric observations are therefore 
potentially much more sensitive 
to emission from the putative faint, outer regions of the torus where
the interaction with the inner regions of the host galaxy must occur.
We suggest that a way to reconcile the evidence for a compact 
torus with these observations and others, such as extended
silicate absorption  (\citet{roche06}, \citet{roche07}) 
and 100 pc-scale CO discs, is that the compact, geometrically and optically 
thick torus is often surrounded by a larger, more diffuse structure, 
associated with the dusty central regions of the host galaxy.  Where the 
torus ends and the host galaxy dust structure starts may be more of a
question of semantics rather than a true physical boundary.  These 
observations examine the interaction between the
host galaxy and possible entrainment into the outer torus regions.
Further multiple-wavelength polarimetric observations of both NGC1068 
and other AGN are required to test this hypothesis.



\acknowledgments

We are grateful to the Gemini, UKIRT and ATC science and engineering staff
for their outstanding work on Michelle and the Gemini telescope, and wish to note especially Chris Carter.  Based on observations obtained at the Gemini Observatory, which is operated by the
Association of Universities for Research in Astronomy, Inc., under a cooperative agreement
with the NSF on behalf of the Gemini partnership: the National Science Foundation (United
States), the Particle Physics and Astronomy Research Council (United Kingdom), the
National Research Council (Canada), CONICYT (Chile), the Australian Research Council
(Australia), CNPq (Brazil) and CONICET (Argentina).




\clearpage

\begin{table}
\begin{center}
\caption{Polarization component summary}
\begin{tabular}{crrr}
\tableline\tableline
Locale, Aperture, $\lambda$ & Degree of Polarization & PA of Polarization & Emitting Component\\
\tableline
Nucleus, 1.7$\arcsec$x1.2$\arcsec$ & 2.48$\pm$0.57\% & 26.7$\pm$15.3$\degr$ & Several\\
9.7$\micron$&&&\\
\tableline
North region & $\sim$2\% & $\sim$8$\degr$ & Dichroic emission from dust\\ 
9.7$\micron$&&&aligned through jet interaction\\
\tableline
East, West, South & $\sim$3.5\% & $\sim$35$\degr$ & Dichroic emission from\\
regions, 9.7$\micron$&&&galactic dust or torus outer edge\\
\tableline
Innermost region & $\le$1\% & - & Dichroic absorption or\\
9.7$\micron$&&&unpolarized source\\
\tableline
Nucleus, 2$\arcsec$ & 1.3$\pm$0.05\% & 49$\pm$3$\degr$ & Several, including dichroic\\
10$\micron$ &&&emission\\
\tableline
Nucleus, 2$\arcsec$ & 4.11$\pm$0.46\% & 120.6$\pm$2.38$\degr$ & Several, including dichroic\\
2.2$\micron$ &&&absorption\\
\tableline
\tableline
\end{tabular}
\tablenotetext{a}{9.7$\micron$ data from this paper}
\tablenotetext{b}{10$\micron$ data from \citet{lums99}}
\tablenotetext{c}{2.2$\micron$ data from \citet{pack97}}
\end{center}
\end{table}

\clearpage

\begin{figure}
\epsscale{.80}
\plotone{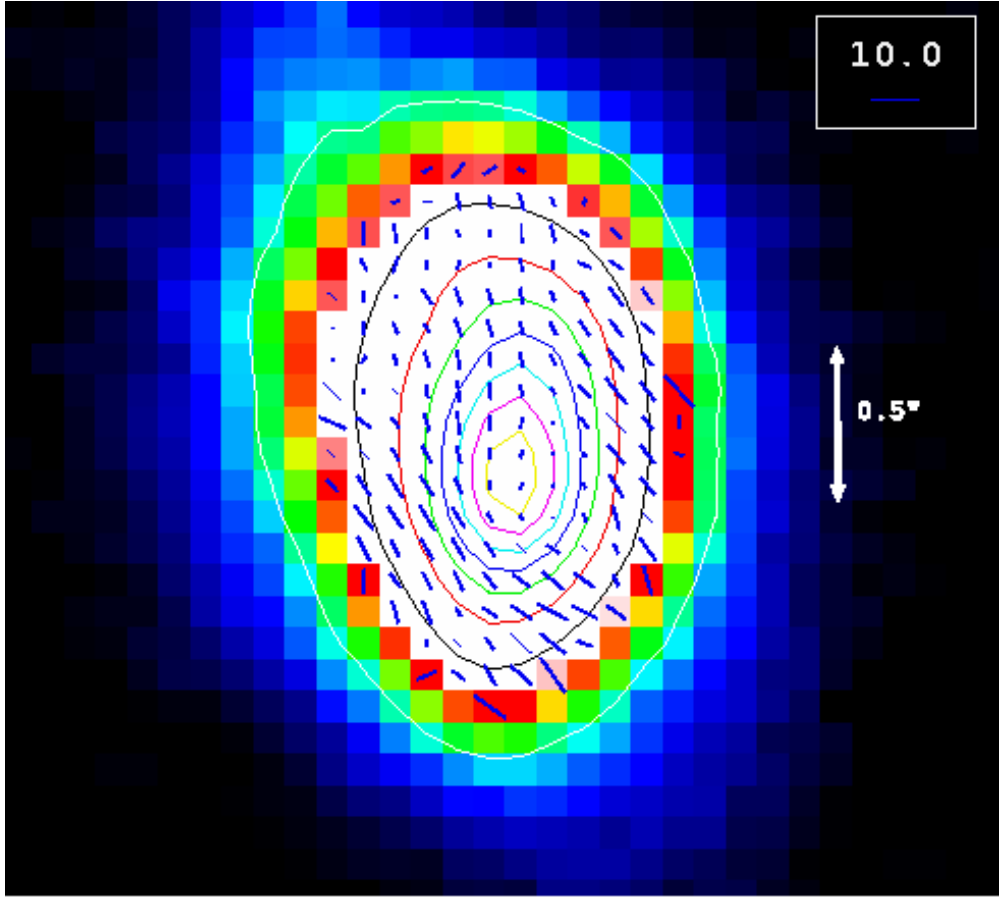}
\caption{Total flux image (color) with the polarization vectors for the
central regions of NGC1068.   The length of the vector is proportional to 
the degree of polarization, and the angle shows the PA of polarization.
Each pixel is 0.1$\arcsec$, and the 10$\%$ polarization scale bar is 
shown in the upper right.  North is up, and east is to the left.}
\label{fig1} 
\end{figure}

\clearpage

\begin{figure}
\plotone{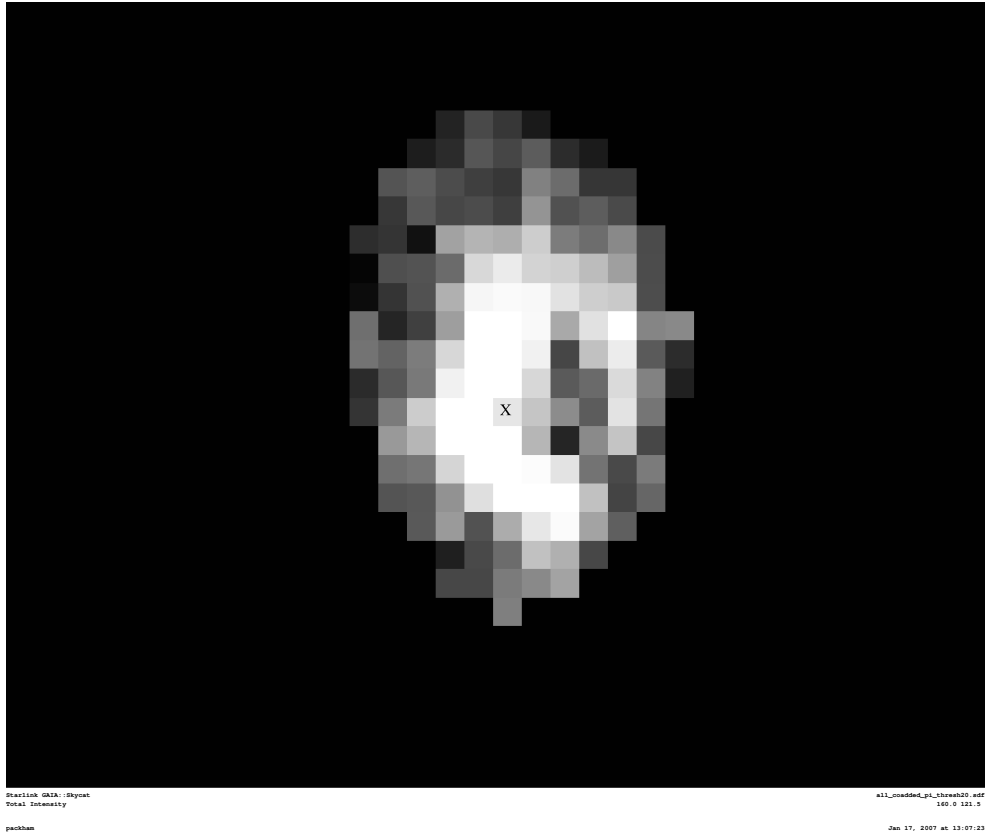}
\caption{Polarized flux image of the central regions of NGC1068.  The
``X'' shows the position of the peak total flux.  Each pixel is 
0.1$\arcsec$. \label{fig2}}
\end{figure}


\begin{thebibliography}{}

\bibitem[Aitken et al.(1984)]{ait84} Aitken, D.~K., Briggs, 
G., Bailey, J.~A., Roche, P.~F., \& Hough, J.~H.\ 1984, \nat, 310, 660 

\bibitem[Antonucci \& Miller(1985)]{ant85} Antonucci, 
R.~R.~J., \& Miller, J.~S.\ 1985, \apj, 297, 621 

\bibitem[Bailey et al.(1988)]{bail88} Bailey, J., Axon, D.~J., 
Hough, J.~H., Ward, M.~J., McLean, I., \& Heathcote, S.~R.\ 1988, \mnras, 
234, 899 

\bibitem[Berry \& Gledhill(2003)]{berry03} Berry, D.~S., 
\& Gledhill, T.~M.\ 2003, Starlink User Notes.  Available from 
http://star-www.rl.ac.uk

\bibitem[Catchpole \& Boksenberg(1997)]{catch97} Catchpole, 
R.~M., \& Boksenberg, A.\ 1997, \apss, 248, 79 

\bibitem[Davies et al.(2006)]{davies06} Davies, R., Genzel, R., 
Tacconi, L., Mueller Sanchez, F., \& Sternberg, A.\ 2006, ArXiv 
Astrophysics e-prints, arXiv:astro-ph/0612009 

\bibitem[Elitzur \& Shlosman(2006)]{elitzur06} Elitzur, M., \& 
Shlosman, I.\ 2006, \apjl, 648, L101 

\bibitem[Galliano et al.(2003)]{galli03} Galliano, E., Alloin, 
D., Granato, G.~L., \& Villar-Mart{\'{\i}}n, M.\ 2003, \aap, 412, 615 

\bibitem[Galliano et al.(2005)]{galli05} Galliano, E., Pantin, 
E., Alloin, D., \& Lagage, P.~O.\ 2005, \mnras, 363, L1 

\bibitem[Gallimore et al.(2001)]{galli01} Gallimore, J.~F., 
Henkel, C., Baum, S.~A., Glass, I.~S., Claussen, M.~J., Prieto, M.~A., \& 
Von Kap-herr, A.\ 2001, \apj, 556, 694 

\bibitem[Glasse et al. (1997)]{glass97} Glasse, A.~C.~H, Atad-Ettedgui,
E.~I., \& Harris, J.~W., SPIE 2871, 1197

\bibitem[Greenhill \& Gwinn(1997)]{green97} Greenhill, L.~J., 
\& Gwinn, C.~R.\ 1997, \apss, 248, 261 

\bibitem[Lumsden et al.(1999)]{lums99} Lumsden, S.~L., Moore, 
T.~J.~T., Smith, C., Fujiyoshi, T., Bland-Hawthorn, J., \& Ward, M.~J.\ 
1999, \mnras, 303, 209 

\bibitem[Jackson et al.(1993)]{jack93} Jackson, J.~M., 
Paglione, T.~A.~D., Ishizuki, S., \& Nguyen-Q-Rieu 1993, \apjl, 418, L13 

\bibitem[Jaffe et al.(2004)]{jaffe04} Jaffe, W., et al.\ 2004, 
\nat, 429, 47 

\bibitem[Lazarian(2003)]{laz03} Lazarian, A.\ 2003, Journal 
of Quantitative Spectroscopy and Radiative Transfer, 79, 881 

\bibitem[Mason et al.(2006)]{mason06} Mason, R.~E., Geballe, 
T.~R., Packham, C., Levenson, N.~A., Elitzur, M., Fisher, R.~S., \& 
Perlman, E.\ 2006, \apj, 640, 612 

\bibitem[Nenkova et al.(2002)]{nen02} Nenkova, M., 
Ivezi{\'c}, {\v Z}., \& Elitzur, M.\ 2002, \apjl, 570, L9 

\bibitem[Ogle et al.(2003)]{ogle03} Ogle, P.~M., Brookings, 
T., Canizares, C.~R., Lee, J.~C., \& Marshall, H.~L.\ 2003, \aap, 402, 849 

\bibitem[Packham et al.(1997)]{pack97} Packham, C., Young, S., 
Hough, J.~H., Axon, D.~J., \& Bailey, J.~A.\ 1997, \mnras, 288, 375 

\bibitem[Packham et al.(2005)]{pack05} Packham, C., Radomski, 
J.~T., Roche, P.~F., Aitken, D.~K., Perlman, E., Alonso-Herrero, A., 
Colina, L., \& Telesco, C.~M.\ 2005, \apjl, 618, L17 

\bibitem[Planesas et al.(1991)]{plan91} Planesas, P., 
Scoville, N., \& Myers, S.~T.\ 1991, \apj, 369, 364 

\bibitem[Radomski et al.(2006)]{radomski06} Radomski, J.~T., et 
al.\ 2006, American Astronomical Society Meeting Abstracts, 209, \#149.11 

\bibitem[Rhee \& Larkin(2006)]{rhee06} Rhee, J.~H., \& Larkin, 
J.~E.\ 2006, \apj, 640, 625 

\bibitem[Roche et al.(1984)]{roche84} Roche, P.~F., Whitmore, 
B., Aitken, D.~K., \& Phillips, M.~M.\ 1984, \mnras, 207, 35 

\bibitem[Roche et al.(2006)]{roche06} Roche, P.~F., Packham, 
C., Telesco, C.~M., Radomski, J.~T., Alonso-Herrero, A., Aitken, D.~K., 
Colina, L., \& Perlman, E.\ 2006, \mnras, 367, 1689

\bibitem[Roche et al.(2007)]{roche07} Roche, P.~F., Packham, 
C., Aitken, D.~K., \& Mason, R.~E.\ 2007, \mnras, 375, 99 

\bibitem[Schinnerer et al.(2000)]{schin00} Schinnerer, E., 
Eckart, A., Tacconi, L.~J., Genzel, R., \& Downes, D.\ 2000, \apj, 533, 850 

\bibitem[Siebenmorgen et al.(2004)]{sieb04} Siebenmorgen, R.,
Kr{\"u}gel, E., \& Spoon, H.~W.~W.\ 2004, \aap, 414, 123 

\bibitem[Simpson et al.(2002)]{simp02} Simpson, J.~P., Colgan, 
S.~W.~J., Erickson, E.~F., Hines, D.~C., Schultz, A.~S.~B., \& Trammell, 
S.~R.\ 2002, \apj, 574, 95 

\bibitem[Smith et al.(2000)]{smith00} Smith, C.~H., Wright, 
C.~M., Aitken, D.~K., Roche, P.~F., \& Hough, J.~H.\ 2000, \mnras, 312, 327 

\bibitem[Watanabe et al.(2003)]{watan03} Watanabe, M., Nagata, 
T., Sato, S., Nakaya, H., \& Hough, J.~H.\ 2003, \apj, 591, 714 

\bibitem[Weigelt et al.(2004)]{weig04} Weigelt, G., 
Wittkowski, M., Balega, Y.~Y., Beckert, T., Duschl, W.~J., Hofmann, K.-H., 
Men'shchikov, A.~B., \& Schertl, D.\ 2004, \aap, 425, 77 

\bibitem[Young et al.(1995)]{young95} Young, S., Hough, J.~H., 
Axon, D.~J., Bailey, J.~A., \& Ward, M.~J.\ 1995, \mnras, 272, 513 

\bibitem[Young et al.(1996)]{young96} Young, S., Packham, C., 
Hough, J.~H., \& Efstathiou, A.\ 1996, \mnras, 283, L1 

\end{thebibliography}
\end{document}